\documentclass[final,5p,times, twocolumn]{elsarticle}

\usepackage{amssymb,amsmath,amsfonts,mathrsfs,bm,gensymb,mathtools}
\usepackage{graphicx,float,subfig}
\usepackage{lineno}

\usepackage{geometry}
\usepackage{hyperref}
\usepackage{tabularx}
\usepackage{array}
\usepackage{color}
\usepackage{booktabs}

\definecolor{gray}{RGB}{170, 170, 170}

\hypersetup{
    colorlinks=true,
    linkcolor=black,
    citecolor=black,
    filecolor=black,
    urlcolor=black,
}
\makeatother

\biboptions{sort&compress}

\begin{document}

\begin{frontmatter}

\title{Tailoring {asymmetry for anisotropic friction} in kirigami metamaterial skins\\with pop-up folding hinges}

\author[add1]{Hamid Reza Tohidvand\fnref{eqc}}
\author[add2]{Alexis White}
\author[add1]{Ali Khosravi\corref{cor1}}
\ead{ali.khosravi@auburn.edu}
\author[add2]{Paolo Celli\fnref{eqc}\corref{cor1}}
\ead{paolo.celli@stonybrook.edu}

\address[add1]{Department of Civil and Environmental Engineering, Auburn University, Auburn, AL 36849, USA}
\address[add2]{Department of Civil Engineering, Stony Brook University, Stony Brook, NY 11794, USA}

\fntext[eqc]{Equal contributors.}
\cortext[cor1]{Corresponding authors.}

\begin{abstract}

Kirigami metamaterial sheets and tubes, owing to their capacity to undergo large elastic deformations while developing three-dimensional surface textures, have enormous potential as skins for soft robots. Here, we propose to use kirigami skins with folding hinges in this same context. These recently-introduced kirigami feature counter-rotating panels connected by pop-up folding hinges. So far, researchers have only explored auxetic and highly-symmetric versions of such patterns. Yet, some of these attributes have to be relaxed in order to explore their full potential as robotic skins. Thus, we parameterize these patterns and relax symmetry constraints, with the goal of using this same platform to obtain a wide range of shape-morphing behaviors.  {We derive kinematic formulas to explore the vast symmetry-enabled design space. We then use numerical simulations and experiments to validate the kinematic predictions and to explore the morphing mechanics of tubular skins. Finally, via experiments, we provide preliminary evidence of the anisotropic friction enabled by patterns with asymmetric pop-ups}. We therefore demonstrate that it is possible to tailor parameters  {in kirigami with folding hinges} to obtain skins that globally expand or contract due to axial elongation, and that present asymmetric pop-ups that yield anisotropic friction --- the most desired attribute for one-way locomotion of soft robots. 

\vspace{10px}
\normalsize{\textbf{This article may be downloaded for personal use only. Any other use requires prior permission of the authors and Elsevier. This article appeared in}: \emph{International Journal of Mechanical Sciences} 296, 110258 (2025) \textbf{and may be found at}: \url{https://doi.org/10.1016/j.ijmecsci.2025.110258}}
\end{abstract}


\end{frontmatter}

\section{Introduction}
\label{s:intro}

In recent decades, mechanical metamaterials have gained increasing attention due to their ability to manifest mechanical properties beyond conventional materials, such as negative Poisson's ratio (auxeticity)~\cite{grima2000auxetic, milton2013complete}, negative stiffness~\cite{jaglinski2007composite, hewage2016double}, or quasi-zero-stiffness~\cite{yan2024double, liang2024design}. These properties stem more from the geometry of the unit cells than from the material they are made of~\cite{bertoldi2017flexible}, and often emerge from the collective behavior of such unit cells. By altering the geometry of these building blocks, designers can completely alter the behavior of metamaterials. Of particular interest to us are metamaterials designed to undergo extreme shape changes and that can morph between distinct flat shapes~\cite{grima2000auxetic, schenk2013geometry}, from flat to  {three dimensional (3D)}~\cite{callens2018flat, celli2018shape} or between distinct 3D shapes~\cite{coulais2016combinatorial, overvelde2017rational, jin2020kirigami, li20213d} when subjected to mechanical stimuli. Such shape-morphing metamaterials have potential applications across scales, from nano- to micro-scale flexible electronic devices~\cite{xu2015assembly, rogers2016origami}, to tabletop-scale soft robots~\cite{rafsanjani2019programming, ze2022soft, dikici2022piece} and to large-scale morphing structures~\cite{melancon2021multistable, zhu2024large, li2024adaptive}.  {Metamaterials can also be designed to exhibit multiple stable configurations with distinct shapes~\cite{rafsanjani2016bistable, li2020theory, wan2024finding, peng2024programming, qiao2024anisotropic} --- an attribute that can be leveraged for energy harvesting, sensing, and robotics applications. Beyond purely mechanical stimuli, researchers are also investigating the use of other triggers to induce shape changes in metamaterials, such as light, heat, and magnetic fields ~\cite{xu2019thermally, taniker2020temperature, lin2024thermal, wu2024heterogeneous, zhang20244d, pal2023programmable}.} 

Among shape-morphing metamaterials, particularly prominent are those that are based on origami and kirigami design principles~\cite{callens2018flat, zhai2021mechanical, tao2023engineering, jin2024engineering, misseroni2024origami}. These involve thin sheets with folds, folds and cuts, or cuts alone. In origami and kirigami with folds and cuts, the presence of such discontinuities allow for the appearance of mechanism modes of deformation (referred to simply as \emph{mechanisms} in the following) that manifest as large, low-energy relative rotations between panels~\cite{schenk2013geometry, filipov2015origami, dorn2022kirigami}. In cut-only kirigami, the extreme morphing capacity can also stem from large relative rotations between panels~\cite{hutchinson2006structural, tang2017design}, or can be ascribed to localized~\cite{isobe2016initial, rafsanjani2017buckling, yang2018multistable, du2023auxetic} or global~\cite{celli2018shape, choi2019programming} buckling.  {Such complex morphing attributes of origami and kirigami metamaterials have made them particularly appealing in bio-inspired engineering. On one hand, such metamaterials can be leveraged to replicate the morphing behavior of natural systems~\cite{hu2023origami, wang2024tensile}; on the other end, some natural systems owe their morphing behavior to origami and kirigami-like designs~\cite{faber2018bioinspired, misseroni2024origami}.}

\begin{figure}[!htb]
\centering
\includegraphics[scale=1]{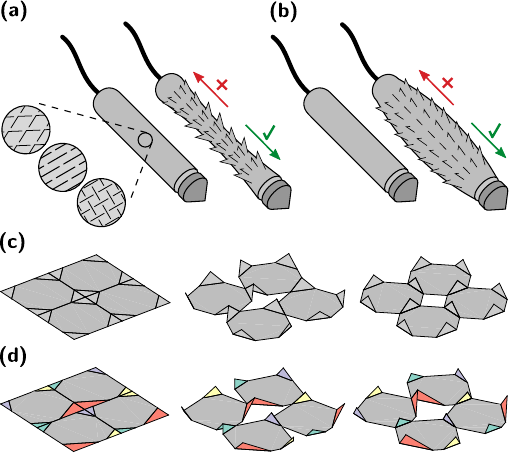}
\caption{Motivation and idea. (a) Soft crawling/burrowing robots rely on kirigami skins to generate anisotropic friction for one-way locomotion (patterns inspired by Ref.~\cite{rafsanjani2019propagation}). (b) We pursue kirigami skins that display anisotropic friction and vastly tailorable morphing behaviors, including auxeticity. (c) Unit cell and kinematic motion of the kirigami with folding hinges by Tang et al.~\cite{tang2019programmable}. (d) Example of unit cell produced in this work, that features asymmetric pop-ups and that can yield anisotropic friction.}
\label{f:idea}
\end{figure}

 {Owing to their attributes and potential in bio-inspiration, origami and kirigami metamaterials are becoming prevalent in soft robotics. For example, robots inspired by origami principles can fold into compact forms for navigation through constrained environments and subsequently unfold to perform more complex tasks~\cite{misseroni2024origami, liu2023orimimetic, zhou2023low}. Origami crawlers and grippers are also extremely popular~\cite{fang2017origami, li2019vacuum, ze2022soft}.} Due to their extreme morphing capacity, kirigami metamaterials have also found applications in various areas of soft robotics, from grasping~\cite{yang2021grasping, hong2022boundary} to actuation~\cite{lipton2018handedness} and locomotion~\cite{rafsanjani2018kirigami, tang2019programmable, sedal2020design, cheng2020kirigami}.  {Locally-buckling kirigami sheets, in particular, are a great platform to achieve directional friction, which is essential for efficient crawling locomotion~\cite{ rafsanjani2018kirigami, seyidouglu2023reconfigurable, parvaresh2024metamaterial}. Specifically, when wrapping such sheets in tubular shapes and when pulling on them axially via a linear actuator, the tailored cuts of the kirigami can produce pop-ups that remain flush in one direction, enabling smooth low-friction movement, while engaging with the surface When displaced in the opposite direction, increasing friction and providing enhanced grip. The resulting directional friction facilitates directed locomotion, promoting forward movement while mitigating backward slippage, thus effectively mimicking the mechanics of crawling organisms such as snakes~\cite{rafsanjani2018kirigami, rafsanjani2019propagation}.}  This concept is
shown in Fig.~\ref{f:idea}(a), where we also illustrate some of the cut patterns that can be used in this context, inspired by Ref.~\cite{rafsanjani2019propagation}. Recent developments in this context have revolved around creating skins for reconfigurable steerability, whereby the same skin can provide anisotropic friction in different directions depending on how it is actuated~\cite{seyidouglu2023reconfigurable}. Other works have explored the applicability of similar morphing principles to burrowing robots, designed to burrow underground following forms akin to those of worm-like organisms~\cite{Liu2019, huang2020effects}. 

When designing kirigami skins for soft crawling robots, researchers have not been very concerned with the radial behavior of the skin, which typically contracts when expanded due to its positive Poisson's ratio. Yet, for underground burrowing and other potential applications like pipe inspection, the radial behavior of the skin is very important, since the robot will have to exert pressure on the surrounding medium to propel itself~\cite{Khosravi2018,Chen2020bio,Martinez2020,Chen2021}. In turn, this would require skins with negative Poisson's ratio, that radially expand when pulled axially. Yet, kirigami architectures that simultaneously present negative Poisson's ratio and asymmetric pop-ups that can yield anisotropic friction, producing qualitative results as shown in Fig.~\ref{f:idea}(b), are not common. One promising pattern in this sense was recently introduced by Tang et al.~\cite{tang2019programmable} and is shown in Fig.~\ref{f:idea}(c). This is a variation of the rotating squares, planar kirigmi introduced by Grima and Evans~\cite{grima2000auxetic}, where the ideal point-like hinges are replaced with initially-flat folding hinges that pop up as the unit cell expands. This same geometry was reprised to study its applicability to elastic wave control~\cite{li2021geometric}, and its out-of-plane folding kinematics~\cite{sempuku2021self}. Yet, the highly symmetric versions of this pattern studied so far lack one of the fundamental attributes we are looking for: asymmetric pop-ups along the actuation direction, as shown in Fig.~\ref{f:idea}(d).  

In this work, we start from the above-mentioned kirigami pattern with folding hinges and rationally explore its design space with the goal of identifying how to tune its parameters to tailor the effective Poisson's ratio and produce asymmetric pop-ups. To do so, we introduce a generic parameterization of the unit cell and explore the relaxation of symmetry constraints within it.  {Our study begins with a kinematic analysis of the unit cell's deformation, to probe the vast design space and understand how folding hinge parameters relate to in-plane kinematics; this analysis yields a few representative symmetric and asymmetric patterns which we subject to further analysis. We perform finite element (FE) simulations of unit cells and experiments aimed at validating the kinematic predictions.  We then fabricate and test tubular skins and to understand their mechanics and the influence of the sheet's thickness, comparing these results to numerical simulations. Finally, we perform experiments aimed at validating our claim on the anisotropic friction of patterns with asymmetric pop-ups. Our results illustrate that it is indeed possible to find combinations of parameters that yield all the desired attributes: tunable Poisson's ratio, asymmetric pop-ups and, consequently, anisotropic friction. Additionally, they highlight the importance of choosing the correct parameters for the tubular skins behave as their planar counterparts.} In summary, our study provides comprehensive information on the morphing behaviors that can be achieved by kirigami with folding hinges, thus opening the way for others to use such patterns in soft robotics or other applications.

This article is organized as follows. In Section~\ref{s:geometry}, we introduce details on the unit cells and their parameters.  {In Section~\ref{s:kinematics}, we perform kinematic studies to explore the vast parameter space of kirigami with folding hinges and identify patterns with desired behaviors.} In Section~\ref{s:validation}, we compare kinematic, numerical and experimental results for some representative patterns.  {Numerical and experimental results on the morphing mechanics of tubular skins, and preliminary results on the anisotropic friction stemming from asymmetric patterns, are reported in Section~\ref{s:tubular}.} Concluding remarks are given in Section~\ref{s:conclusion}.

\begin{figure*}[!htb]
\centering
\includegraphics[scale=1]{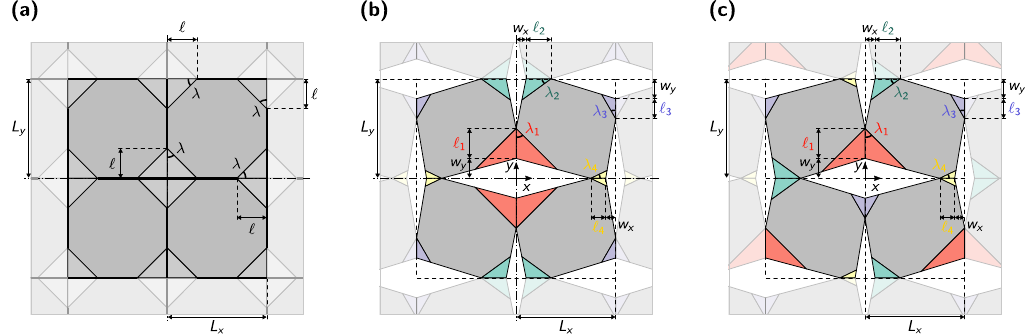}
\caption{Families of unit cell geometries and design parameters. (a) Unit cell geometry as introduced by Tang et al.~\cite{tang2019programmable}, in which the only design parameters are representative lengths $L_x$ and $L_y$, fold length $\ell$ and fold angle $\lambda$; here the thicker lines represent cuts/slits. This geometry displays two mirror symmetry axes and, when actuated, evolves as shown in Fig.~\ref{f:idea}(c). (b) Symmetric variation of the unit cell in (a), where we replace slits with diamond-shaped cuts with half-diagonals $w_x$ and $w_y$, and we allow for parameter variations across pop-ups ($\lambda_i$ and $\ell_i$, with $i=1, \ldots 4$, can be chosen independently), while maintaining mirror symmetries. (c) Asymmetric version of the unit cell in (b), where the pop-up labeling has changed as to potentially break mirror symmetries.}
\label{f:geometry}
\end{figure*}

\section{Unit cell geometry and design parameters}
\label{s:geometry}

The kirigami unit with folding hinges by Tang et al.~\cite{tang2019programmable} is shown in Fig.~\ref{f:geometry}(a). It is composed of four octagonal panels, each one inscribed within a square of size $L_x$ by $L_y$. It is assumed that the unit needs to be tiled periodically in space to create a larger pattern; thus, lattice vectors have magnitude of $2L_x$ and $2L_y$ along the horizontal and vertical directions, respectively. Each panel is connected to parts of its neighbors via folding hinges, here represented as lighter-gray triangles whose dimensions can all be obtained from their length $\ell$ and angle $\lambda$. Slits between panels, here shown as thicker lines, guarantee that the unit cell deforms according to the desired mechanism, featuring in-plane counter-rotations of the panels and pop-up of the hinges, with all parts remaining rigid and deformations being concentrated at ideal fold lines as illustrated in Fig.~\ref{f:idea}(c). This kirigami geometry possesses three distinct traits: it is symmetric about vertical and horizontal axes passing through the center of the cell; it has negative Poisson's ratio; and it features identical folding hinges that all pop up of the same amount.

The first modification we introduce is to depart from line-like slits and replace them with diamond-shaped cuts. This echoes modifications made in the past to the rotating-squares pattern, that allowed to tune its effective Poisson's ratio across a wide range of positive and negative values~\cite{celli2018shape, zheng2022continuum}. As shown in Figs.~\ref{f:geometry}(b), (c), each slit is either defined by its horizontal half-diagonal, $w_x$, or by its vertical one, $w_y$; its other dimension follows from its nearest neighboring cuts and from the folding hinges. We here choose to define the central slit of the cell by $w_y$, its closest neighbors to the left, right, top and bottom by $w_x$, and the diagonally closest ones again by $w_y$ and so on. 

The second modification we apply is to relax the constraint that all folding hinges are identical. There are two ways we can do this. One way entails considering all hinges in the top-right quadrant of each unit cell different from each other, and to apply mirror symmetry about the Cartesian axes $x$ and $y$, to obtain a pattern as shown in Fig.~\ref{f:geometry}(b) --- labeled as \emph{symmetric} in the following. Here and in the following, we color code hinges according to their type. We then define a hinge length parameter $\ell_i$ and an angle parameter $\lambda_i$ to characterize each hinge type, with $i=1, \ldots 4$. It is useful to point out that patterns as in Fig.~\ref{f:geometry}(b) are symmetric for all parameter choices.

The third and final modification we introduce involves relaxing symmetry constraints about the $x$ and $y$ axes. We again consider the top-right quadrant of the cell and assign different parameters to all of its folding hinges. Then, we assign parameters to the hinges of the neighboring panels following these rules:  each hinge needs to have the same properties in all its portions; hinges that face each other across slits have to be of different types; and hinges have to be assigned so that periodicity is preserved while keeping the unit cell size to 2 $\times$ 2 panels. What results from this process are designs as the one in Fig.~\ref{f:geometry}(c) --- called \emph{asymmetric} in the following --- where any symmetry about the $x$ and $y$ axes is lost for most parameter choices, while their ``tileability'' in space to produce periodic patterns is preserved. These designs are the only ones that hold promise in the context of robotic skins with frictional anisotropy: if, for example, we consider the geometry in Fig.~\ref{f:geometry}(c) and imagine using it for a robot that expands along the $y$ axis, then the fact that pop-ups of type 1 are larger than those of type 3 will likely cause the robot to move  {differently along $+y$ and $-y$ directions}. 

Throughout this work, we set the panel dimensions equal to each other: $L_x=L_y$. Additionally, the characteristic dimensions of the cuts and the folding hinge lengths are normalized with respect to the panel dimension along the direction they are aligned with: $\bar{w}_x=w_x/L_x$, $\bar{w}_y=w_y/L_y$, $\bar{\ell}_1=\ell_1/L_y$, $\bar{\ell}_2=\ell_2/L_x$, $\bar{\ell}_3=\ell_3/L_y$ and $\bar{\ell}_4=\ell_4/L_x$.

\section{Kinematic analysis and parametric exploration}
\label{s:kinematics}

 {We now try to navigate the vast design space introduced in the previous section, with the goal of understanding which combinations of parameters preserve the \emph{global mechanism} featuring panel counter-rotations, and identifying geometries that yield asymmetric pop-ups. We are particularly interested in understanding how the pop-up parameters relate to in-plane kinematics --- an aspect that is unique to kirigami metamaterials with folding hinges.}

 {As previously mentioned, the metamaterials we investigate here are variations of the rotating squares pattern. In ~\ref{a:mechanisms}, by analyzing the mechanisms and states of self stress of such patterns, we show that the presence of folding hinges does not affect the global mechanism. We also show that, in the case of asymmetric geometries, we cannot choose hinge lengths independently (we need $\bar{\ell}_1=\bar{\ell}_3$ and $\bar{\ell}_2=\bar{\ell}_4$), while hinge angles can be chosen independently. These considerations considerably restrict the available design space; however, the possibility of choosing hinge angles independently can yield interesting asymmetric designs, as shown below. Our conclusions echo recent findings on the need of having parallelogram-shaped slits to preserve the global mechanism of rotating squares metamaterials and their variations~\cite{yang2018geometry, singh2021design, zheng2022continuum}, including a formal theorem on this matter by Dang et al.\ that extends to patterns with even less symmetry than ours~\cite{dang2021theorem}.}

\subsection{Kinematic analysis}

 {We now probe this restricted design space} to identify geometries that yield asymmetric pop-ups. This parametric study is guided by nonlinear kinematic formulas, that characterize the finite-amplitude mechanism motion of the selected unit cells. We begin by introducing our kinematic formulas for a generic unit cell and continue by performing the parametric study and identifying representative patterns to be further analyzed.

\begin{figure*}[!htb]
\centering
\includegraphics[scale=1]{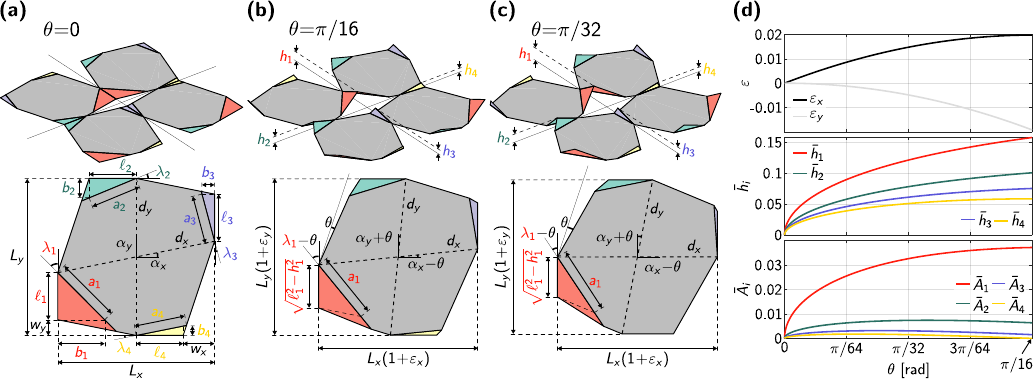}
\caption{Example of kinematic simulation. (a)-(c) Three instances of the kinematic evolution of a asymmetric unit cell with mechanism, obtained by setting slit openings $\bar{w}_x=0.2$, $\bar{w}_y=0.1$, fold lengths $\bar{\ell}_1=\bar{\ell}_2=\bar{\ell}_3=\bar{\ell}_4=0.3$ and fold angles $\lambda_1=\pi/4$, $\lambda_2=\pi/8$, $\lambda_3=\pi/12$, $\lambda_4=\pi/16$. The panel rotation angle $\theta$ is the only DOF; specifically $\theta=0$ in (a), $\theta=\pi/16$ in (b) and $\theta=\pi/32$ in (c). (d) Evolution of strains, normalized heights of the pop-ups and normalized opening areas of the pop-ups as a function of $\theta$, for the unit cell in (a).}
\label{f:desparam}
\end{figure*}

To explain how we derive our kinematic formulas, we consider the pattern shown in Fig.~\ref{f:desparam}(a). In particular, we choose all hinge lengths to be identical to preserve the global mechanism. To be as generic as possible, we select different values of slit opening $\bar{w}_x$ and $\bar{w}_y$, and we choose distinct values for all hinge angles $\lambda_i$. Two instances of the kinematic evolution of the selected unit cell as a function of the degree of freedom (DOF) $\theta$ are shown in Figs.~\ref{f:desparam}(b), (c). Our kinematic formulas are obtained  {by tracking the motion of a single panel, and we choose the top-right one.} 

To track the overall strains of the unit in the $x$ and $y$ directions, we draw two dashed panel diagonals connecting opposite rotation points of the panel, i.e., connecting the tip of hinge 1 to that of hinge 3, and the tip of hinge 2 to that of hinge 4, as shown in Fig.~\ref{f:desparam}A. The lengths of these diagonals are labeled as $d_x$ and $d_y$, respectively, and can be defined with respect to the other geometrical parameters of the unit, as follows:
\begin{equation}
    \begin{split}
        d_x=\sqrt{L_x^2+( L_y-2w_y-\ell_1-\ell_3 )^2},\\
        d_y=\sqrt{L_y^2+( 2w_x+\ell_2+\ell_4-L_x )^2}.
    \end{split}
    \label{e:diag}
\end{equation}
The initial angle of $d_x$ with the $x$ axis is labeled $\alpha_x$ and the angle between $d_y$ and $y$ is $\alpha_y$; these angles can be computed as
\begin{equation}
    \begin{split}
        \alpha_x=\arctan \left( \frac{L_y-2w_y-\ell_1-\ell_3}{L_x} \right),\\
        \alpha_y=\arctan \left( \frac{2w_x+\ell_2+\ell_4-L_x}{L_y} \right).
    \end{split}
    \label{e:alph}
\end{equation}
As the panels in the unit counter-rotate by angle $\theta$, the above mentioned diagonals in the top-right panel rotate so that the angle between the diagonal of length $d_x$ and $x$ becomes $\alpha_x-\theta$, while the angle between $d_y$ and $y$ becomes $\alpha_y+\theta$. The projection of the diagonal of length $d_x$ onto the $x$ axis and that of $d_y$ onto $y$ can then be used to find the overall strains of the cell,
\begin{equation}
    \begin{split}
        \varepsilon_x=\frac{d_x \cos{(\alpha_x-\theta)} -L_x}{L_x},\\
        \varepsilon_y=\frac{d_y \cos{(\alpha_y+\theta)} -L_y}{L_y}.
    \end{split}
    \label{e:strain}
\end{equation}
From these formulas, it is possible to see that hinge angles do not play a role on the overall kinematics of the unit cell; this further reinforces the validity of our conjecture on the independence of the global mechanism from the local mechanisms of hinge pop-up. 

The evolution of $\varepsilon_x$ and $\varepsilon_y$ for the unit in Fig.~\ref{f:desparam}(a) as a function of $\theta$ is shown in the top panel of Fig.~\ref{f:desparam}(d). We only consider a range of $\theta$ between 0 and the smallest hinge angle ($\lambda_4$ here), to avoid overlap between panels. In this specific case, we can see that the horizontal and vertical strains increase and decrease, respectively, in a monotonic fashion, yielding a positive effective Poisson's ratio ($\nu=-\varepsilon_y/\varepsilon_x$) throughout the kinematic evolution of the cell. Please note that these strains need not be monotonic, as the unit cell can expand and then contract in any direction as $\theta$ increases. In the following, though, we will further restrict the range of $\theta$ for each cell such that $\varepsilon_x$ is only positive.

To track the evolution of the pop-ups with $\theta$, we derive formulas for their heights.  {The steps of this derivation are shown in \ref{a:kin}. The final result is}
\begin{equation}
        h_i=\ell_i \sqrt{\frac{cos(2 \lambda_i - 2\theta) - cos(2 \lambda_i) }{cos(2\lambda_i - 2\theta) + 1}}.
    \label{e:height}
\end{equation}
Provided that all panels within the unit rotate of the same amount, which is a requirement to have the desired global mechanism, this equation can be applied to determine the height of any pop-up, independently of where it is located within the unit. In the following, pop-up heights are normalized as follows: $\bar{\ell}_1=\ell_1/L_y$, $\bar{\ell}_2=\ell_2/L_x$, $\bar{\ell}_3=\ell_3/L_y$, $\bar{\ell}_4=\ell_4/L_x$. The evolution of the pop-up heights with $\theta$ for the example geometry is shown in the middle panel of Fig.~\ref{f:desparam}(d). In general, these curves always increase; while they start steep, they tend to flatten as $\theta$ reaches its limit.

Finally, we introduce another quantity, that better characterizes the asymmetry of the pop-up hinges. We define the pop-up opening area $A_i$ (normalized as $\bar{A}_i=A_i/(L_x L_y)$) as the triangular area with height equal to the pop-up height $h_i$ and base equal to the projection of the segment of length $b_i$ onto the $x$-$y$ plane and then onto the $x$ axis:
\begin{equation}
        A_i=h_i a_i \sin{(\lambda_i-\theta)},
    \label{e:area}
\end{equation}
 {where $a_i$ and $b_i$ are defined in Fig.~\ref{f:desparam}(a) and in \ref{a:kin}.}
This area allows monitoring of the relative opening between pop-ups ---  {an aspect that, together with the pop-up height and the dimensions of the folding hinges, is bound to have an influence on the frictional properties of the kirigami skin}.

Finally, we use these formulas to perform kinematic simulations, that allow us to visualize the evolution of the unit cell with $\theta$. Figs.~\ref{f:desparam}(a)-(c) represent three instances of one such simulation. To obtain these snapshots, we apply planar rotations to the octagonal panels; the pop-ups, on the other hand, are obtained by using the pop-up height equation (Eq.~\ref{e:height}), and by connecting the pop-up peaks to the proper points on the octagons. 

 {Even if we restrict ourselves to patterns with $\ell_1=\ell_3$ and $\ell_2=\ell_4$, not all combinations of the remaining parameters are valid. In fact, some combinations yield unwanted overlaps between panels and hinges. To avoid overlaps, we set limits on these parameters, as discussed in Fig.~\ref{a:kin}}.

\begin{figure*}[!htb]
\centering
\includegraphics[scale=0.84]{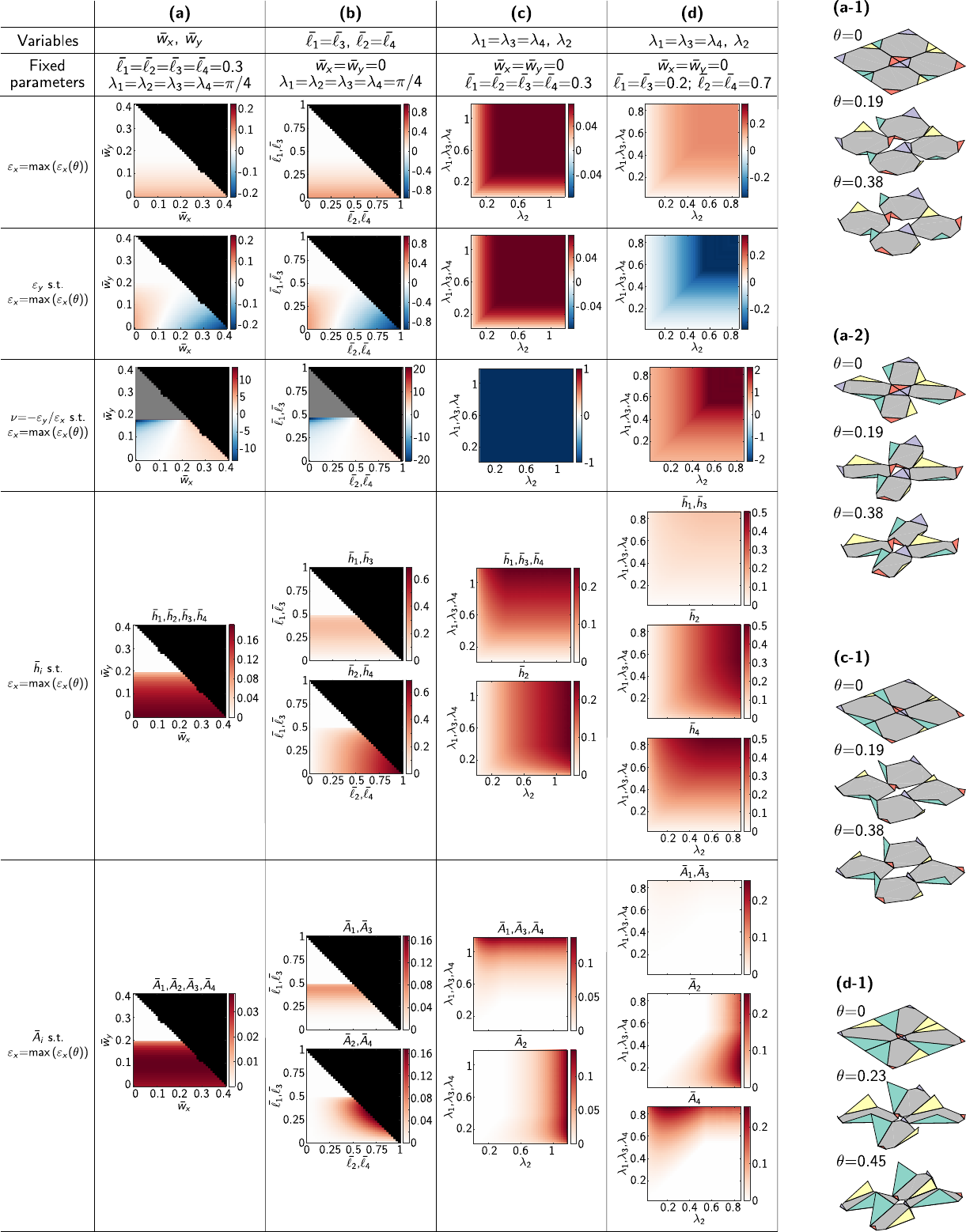}
\caption{ Kinematics-guided design space exploration. Studies (a), (b), (c), (d) are described in detail in the text. (a-1), (a-2) Kinematic simulations of sample geometries with parameters corresponding to the (a) study, and with $\bar{w}_x=\bar{w}_y=0$ and $\bar{w}_x=0.3$, $\bar{w}_y=0$, respectively. (c-1) Kinematic simulation of geometry with parameters corresponding to the (c) study, and with $\lambda_1=\lambda_3=\lambda_4=0.39$ and $\lambda_2=1.15$. (d-1) Kinematic simulation of sample geometry with parameters corresponding to the (d) study, and with $\lambda_1=\lambda_3=\lambda_4=0.45$, $\lambda_2=0.85$. The geometrical parameters for these sample geometries are also listed in Table~\ref{t:rep}.}
\label{f:kinparam}
\end{figure*}

\subsection{Parametric study}
We now have all the tools to study the influence of design parameters on the kinematic response of kirigami unit cells with folding hinges. To summarize the behavior of the unit for each combination of parameters, we consider a snapshot of it at the $\theta$ angle that yields maximum positive strain $\varepsilon_x$; this represents the instance of maximum purely-mechanistic horizontal stretch in a scenario where the unit cell is being pulled along the horizontal direction. Rather than analyzing all parameters in all possible ranges, we perform a limited set of parametric studies with the objective of unveiling some of the effects of such parameters. Results are reported as design maps of the maximum positive strain, $\max (\varepsilon_x(\theta))$, the corresponding strain in the tangential direction, $\varepsilon_y$, the effective Poisson's ratio, $\nu=-\varepsilon_y/\varepsilon_x$, the pop-up heights $h_i$ and opening areas $A_i$ at that same instance of deformation, all as a function of selected sets of parameters. If the combinations of parameters corresponding to a point in a design map yield designs with panel overlaps, that same point is colored black to highlight the inaccessibility of that parameter combination. Gray points in the $\nu$ colormap indicate regions where that quantity is undefined. 

In the study  of Fig.~\ref{f:kinparam}(a), we probe the effects of the slit openings $\bar{w}_x$ and $\bar{w}_y$, while fixing hinge lengths to $\bar{\ell}_1=\bar{\ell}_2=\bar{\ell}_3=\bar{\ell}_4=0.3$ and angles to $\lambda_1=\lambda_2=\lambda_3=\lambda_4=\pi/4$. In the $\varepsilon_x$ plot, we notice that a large portion of the design space, the black region, is not accessible. In the accessible region, we can see that patterns with $\bar{w}_y>0.17$ do not produce any positive strain and only contract along $x$ for all admissible values of $\theta$; these designs are of no interest to us here. We can also see that $\bar{w}_x$ does not affect the maximum strain along $x$. From the $\varepsilon_y$ strain and Poisson's ratio plots (evaluated at the $\theta$ value of maximum positive $\varepsilon_x$), we can see that the slit opening parameters allow to vastly tune the Poisson's ratio of the pattern. In particular, low and high $\bar{w}_x$ values yield negative and positive Poisson's ratios, respectively. For this choice of parameters, the folding hinges are all identical. Thus, all pop-up heights and opening areas behave identically. We can see that the pop-up height correlates with $\varepsilon_x$, while the opening area has a maximum for values of $\bar{w}_y$ around 0.1. To conclude this first study, we select two representative unit cells corresponding to combinations of parameters that yield symmetric designs with negative and positive Poisson's ratios. Their kinematic evolution, for $\theta$ values until the maximum $\varepsilon_x$ is reached, is shown in Figs.~\ref{f:kinparam}(a-1) and (a-2).

In the study Fig.~\ref{f:kinparam}(b), we instead investigate the effects of the hinge lengths. Owing to the previously-determined constraint on hinge lengths, we select as independent parameters $\bar{\ell}_1=\bar{\ell}_3$ and $\bar{\ell}_2=\bar{\ell}_4$, while setting slit openings $\bar{w}_x=\bar{w}_y=0$ and fold angles $\lambda_1=\lambda_2=\lambda_3=\lambda_4=\pi/4$. The $\varepsilon_x$, $\varepsilon_y$ and $\nu$ plots show similar trends with respect to the corresponding figures of the Fig.~\ref{f:kinparam}(a) study, highlighting that hinge lengths can also be used to tune the Poisson's ratio. More interesting here are the maps for pop-up heights and opening areas. These maps show that the height of pop-ups 2 and 4 can be maximized by selecting large lengths for these hinges and small lengths for the others. On the other hand, to maximize the opening areas of pop-ups 2 and 4, one needs to choose $\bar{\ell}_1=\bar{\ell}_3\approx 0.3$ and $\bar{\ell}_2=\bar{\ell}_4\approx 0.7$ --- something that would have been hard to guess without a systematic parametric study.

In the studies of Figs.~\ref{f:kinparam}(c) and (d), we instead concentrate on the effects of the hinge angles, with an eye towards obtaining patterns with drastically-asymmetric pop-ups. We therefore choose as independent parameters hinge angles $\lambda_2$ and $\lambda_1=\lambda_2=\lambda_3$ --- something that would cause the resulting patterns to have frictional asymmetry along the $x$ axis. In the study of Fig.~\ref{f:kinparam}(c), we concentrate on an asymmetric pattern with negative Poisson's ratio. The effects of hinge angles on the extensibility of the cell are minimal, as long as these angles are not so small to cause a premature, complete folding of the hinges. The colormaps for pop-up heights and opening areas show that the region that yields more asymmetry in favor of the pop-up of the 2 type corresponds to large $\lambda_2$ and to small values of all other hinge angles. Thus, we choose the geometry shown in Fig.~\ref{f:kinparam}(c-1), with $\lambda_1=\lambda_3=\lambda_4=0.39$ and $\lambda_2=1.15$, as the representative asymmetric one with negative Poisson's ratio; we did not choose a smaller value for $\lambda_1$ since this could have limited the overall extensibility of the unit. As illustrated by this kinematic simulation, the pop-up of the 2 type is much larger and much more open than the others at the maximum strain level --- which is exactly what we are looking for. In the study of Fig.~\ref{f:kinparam}(d), we instead concentrate on obtaining similar effects in geometries with positive Poisson's ratio with slit openings $\bar{w}_x=\bar{w}_y=0$ and fold lengths $\bar{\ell}_1=\bar{\ell}_3=0.2$ and $\bar{\ell}_2=\bar{\ell}_4=0.7$  {--- values that were close to maximizing the opening area in the study of Fig.~\ref{f:kinparam}(b)}. From the opening area maps, we can see that the maximum asymmetry in pop-up opening areas is achieved with $\lambda_2 = 0.85$ and all other angles $\approx 0.2$. Yet, the strain plots show that there is a tradeoff between asymmetry and maximum cell extension. Thus, as representative asymmetric geometry with positive Poisson's ratio, we select one with $\lambda_1=\lambda_3=\lambda_4=0.45$, $\lambda_2=0.85$, and show its kinematic evolution in Fig.~\ref{f:kinparam}(d-1). This simulation allows us to appreciate that designs with positive $\nu$ can produce much more pop-up asymmetry with respect to those with negative $\nu$.

\section{Response of representative {unit cells}}
\label{s:validation}
 {We now analyze the response of a few representative symmetric and asymmetric unit cells identified in our design exploration. Our analytical kinematics results for these patterns are validated via experiments and numerical FE simulations. For comparison}, we also simulate the behavior of a geometry that, according to our previous analysis, does not feature a global mechanism. The selected representative patterns are listed in Table~\ref{t:rep}, where we indicate the figure(s) they appear in and their design parameters.
\begin{table}[!htb]
\centering\footnotesize
\begin{tabular}{p{0.022\textwidth}p{0.065\textwidth}p{0.007\textwidth}p{0.007\textwidth}p{0.008\textwidth}p{0.008\textwidth}p{0.008\textwidth}p{0.008\textwidth}p{0.014\textwidth}p{0.014\textwidth}p{0.014\textwidth}p{0.02\textwidth}}
    \toprule
       Label & Figures & $\bar{w}_x$ & $\bar{w}_y$ & $\bar{\ell}_1$ & $\bar{\ell}_2$ & $\bar{\ell}_3$ & $\bar{\ell}_4$ & $\lambda_1$ & $\lambda_2$ & $\lambda_3$ & $\lambda_4$ \\
     \midrule
     SN & Fig.~\ref{f:kinparam}(a-1) & 0 & 0 & 0.3 & 0.3 & 0.3 & 0.3 & $\pi/4$ & $\pi/4$ & $\pi/4$ & $\pi/4$ \\
     & Fig.~\ref{f:planar}(a) & & & & & & & & & \\
     \midrule
     SP & Fig.~\ref{f:kinparam}(a-2) & 0.3 & 0 & 0.3 & 0.3 & 0.3 & 0.3 & $\pi/4$ & $\pi/4$ & $\pi/4$ & $\pi/4$ \\
     & Fig.~\ref{f:planar}(b) & & & & & & & & & \\
     \midrule
     AN & Fig.~\ref{f:kinparam}(c-1) & 0 & 0 & 0.3 & 0.3 & 0.3 & 0.3 & 0.39 & 1.15 & 0.39 & 0.39 \\
     & Fig.~\ref{f:planar}(c) & & & & & & & & & \\
     \midrule
     AP & Fig.~\ref{f:kinparam}(c-2) & 0 & 0 & 0.2 & 0.7 & 0.2 & 0.7 & 0.45 & 0.85 & 0.45 & 0.45 \\
     & Fig.~\ref{f:planar}(d) & & & & & & & & & \\
     \midrule
     NM & Fig.~\ref{f:planar}(e) & 0 & 0 & 0.2 & 0.4 & 0.2 & 0.2 & $\pi/6$ & $\pi/4$ & $\pi/6$ & $\pi/6$ \\
     \toprule
\end{tabular}
\caption{Geometrical parameters of the representative designs we select for further analysis, indicating the figures where such patterns appear,  {and some labels, which can be interpreted as follows. SN: symmetric with negative Poisson's ratio; SP: symmetric with positive Poisson's ratio; AN: asymmetric with negative Poisson's ratio; AP: asymmetric with positive Poisson's ratio; NM: no mechanism.}}
\label{t:rep}
\end{table}
 {Details on the fabrication procedures and specimen dimensions, experimental setups and numerical models are discussed in \ref{a:fab}, \ref{a:exp} and \ref{a:num}, respectively.}

\begin{figure*}[!htb]
\centering
\includegraphics[scale=1]{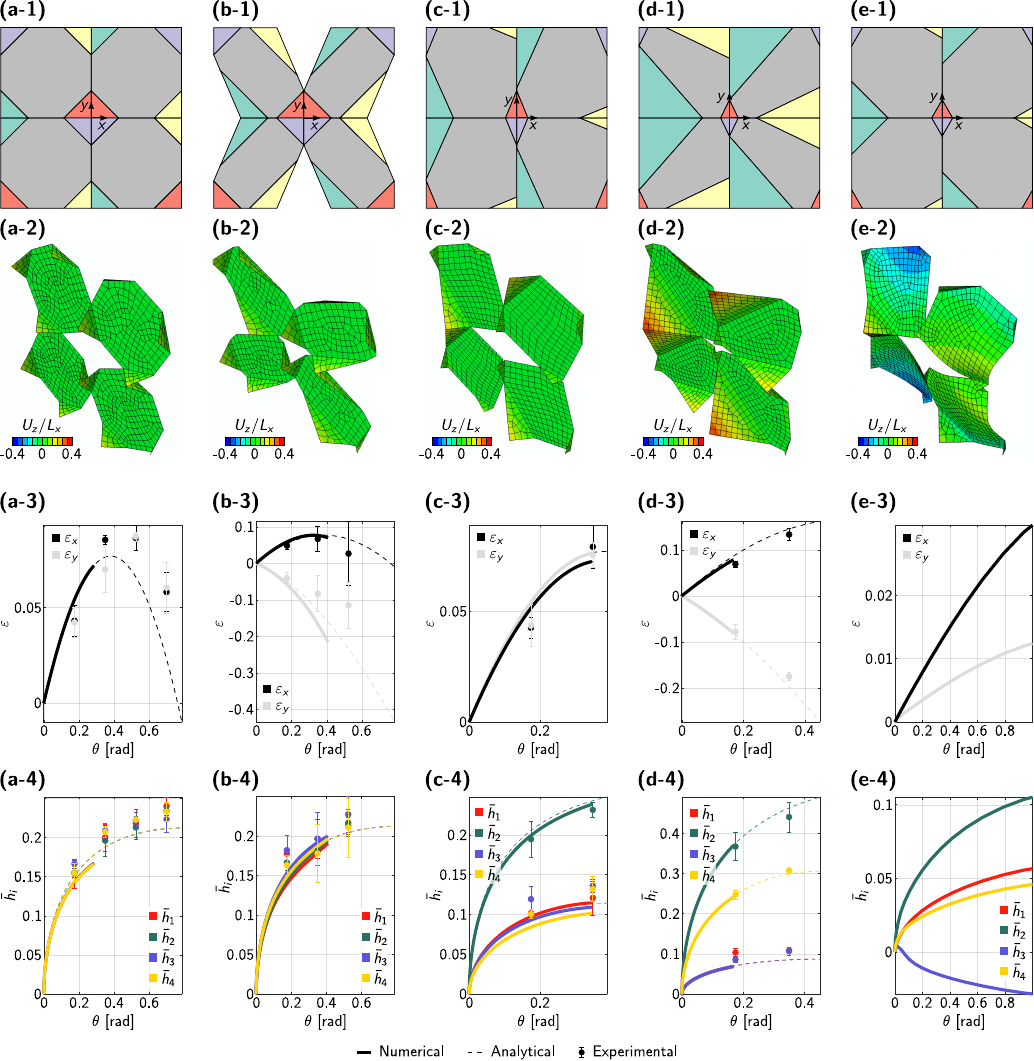}
\caption{ {Comparison between analytical, numerical and experimental response of representative unit cells, labeled as in Table~\ref{t:rep}. (a) Response of the SN design, symmetric with negative Poisson's ratio. (b) Response of the SP design, symmetric with positive Poisson's ratio. (c) Response of the AN design, asymmetric with negative Poisson's ratio. (d) Response of the AP design, asymmetric with positive Poisson's ratio. (e) Response of the NM design, that does not feature a mechanism.} Throughout the figure, ( -1) subfigures show the geometry, ( -2) show a numerically-simulated snapshot of the deformed structure, ( -3) and ( -4) show a comparison between analytical, numerical and experimental results for the strains and pop-up heights, respectively, as a function of the actuation angle $\theta$, while ( -5) show the stress-strain numerical response. Experimental results represent the mean and standard deviation of three measurements.}
\label{f:planar}
\end{figure*}

We begin the analysis of the patterns listed in Table~\ref{t:rep} by studying the auxetic one with symmetric pop-ups shown in Fig.~\ref{f:planar}(a-1) (akin to the one by Tang et al.~\cite{tang2019programmable}). Fig.~\ref{f:planar}(a-2) illustrates a snapshot of the FE simulation of the same unit, which allows to see the mesh used for its analysis. Quantitative results are shown in Figs.~\ref{f:planar}(a-3), (a-4). From the plot of strains vs.\ $\theta$ in Fig.~\ref{f:planar} (a-3), we can see that numerical simulations and analytical calculations overlap perfectly, and that the experimental data points follow a very similar up-down trend. The pop-up height plots in Fig.~\ref{f:planar}(a-4) also show a  remarkable consistency with experiments. Small discrepancies with the numerical simulations are due to the fact that, due to periodic boundary conditions, we do not constrain the out-of-plane displacement of the octagonal panels; thus, small out-of-plane deflections can alter the pop-up heights. We now turn to the pattern with symmetric pop-ups  {and positive Poisson's ratio of Figs.~\ref{f:planar}(b-1), (b-2). The quantitative results in Figs.~\ref{f:planar}(b-3), (b-4) show again consistency of trends between analytics, experiments and numerics.} We now move to the analysis of patterns with asymmetric pop-ups, starting from the one in Fig.~\ref{f:planar}(c-1), whose FE model at some level of deformation is shown in Fig.~\ref{f:planar}(c-2).  {The strain evolution of Fig.~\ref{f:planar}(c-3) and the pop-up height plots in Fig.~\ref{f:planar}(c-4) show again good consistency between experiments and models. Similar considerations can be made for the non-auxetic pattern with asymmetric pop-ups shown and described in Figs.~\ref{f:planar}(d-1)-(d-4). }

Finally, we resort to numerical simulations alone to study the pattern in Figs.~\ref{f:planar}(e-1), (e-2); owing to the fact that ${\ell}_2$ is different from all other hinge lengths, this pattern does not feature a mechanism and therefore cannot be studied with our analytical kinematics tools, and its morphing attributes cannot be easily quantified in experiments. The strain plot in Fig.~\ref{f:planar}(e-3) shows that this pattern has mild auxeticity, but less than the similar pattern with mechanism of Fig.~\ref{f:planar}(c-1). This is due to the fact that, as shown in Fig.~\ref{f:planar}(e-2), the panels undergo significant bending. The absence of mechanisms also causes the pop-ups to be all different from each other, as shown in Fig.~\ref{f:planar}(e-4).  

\begin{figure*}[!htb]
\centering
\includegraphics[scale=1]{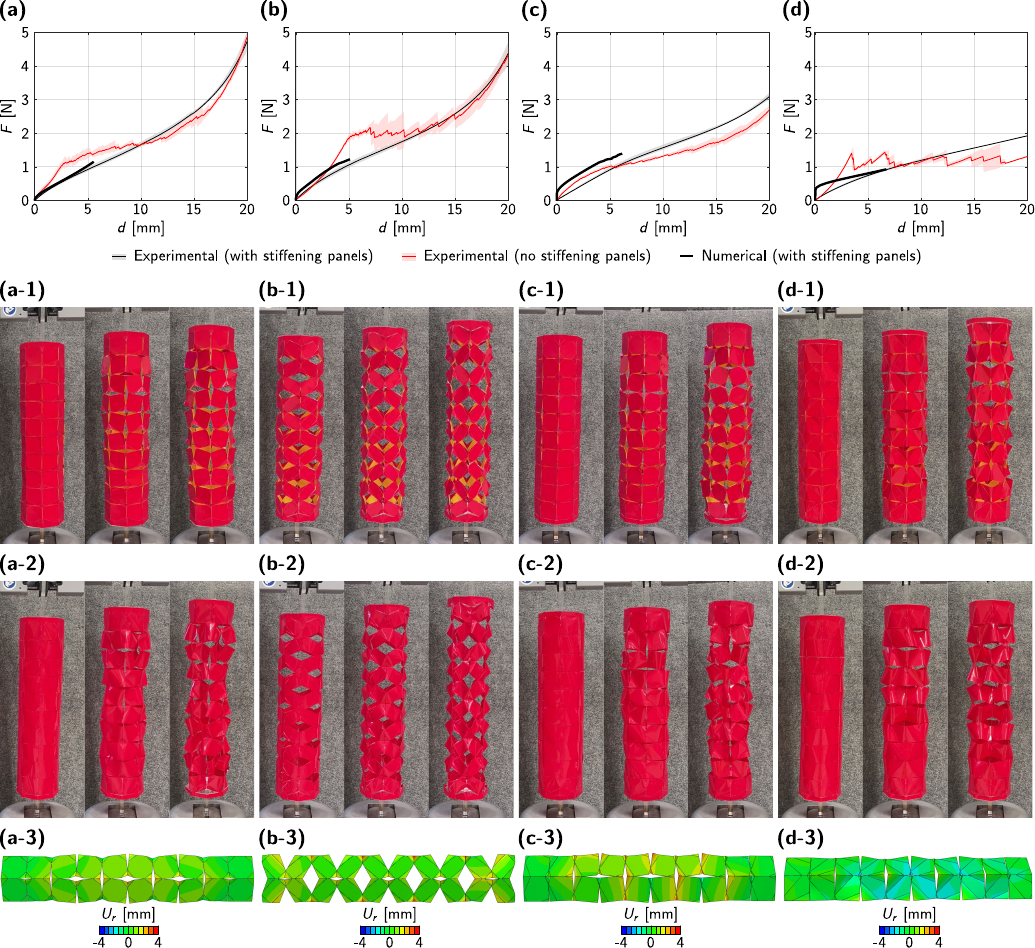}
\caption{Results on the morphing mechanics of tubular structures with different geometrical designs, labeled as in Table~\ref{t:rep}, with and without stiffening panels; experimental results are compared to numerical simulations (only for the cases with stiffening panels). (a) Response of the SN design, symmetric with negative Poisson's ratio. (b) Response of the SP design, symmetric with positive Poisson's ratio. (c) Response of the AN design, asymmetric with negative Poisson's ratio. (d) Response of the AP design, asymmetric with positive Poisson's ratio. Throughout the figure, ( -1) and ( -2) subfigures show experimental snapshots of the deformation of tubes with and without stiffening panels, respectively; ( -3) subfigures show numerical snapshots of the deformation of tubes with panels, where the colormap indicates radial displacement.}
\label{f:tubes}
\end{figure*}

{
\section{From units to tubular skins}
\label{s:tubular}

In this section, we move from single units to tubular kirigami skins. We first perform experiments and simulations on the morphing mechanics of the tubes; we then perform preliminary experiments on the frictional properties of the skins when dragged along a rough surface, with the goal of evaluating the presence/absence of anisotropic friction. The tubes we fabricate and test are characterized by the mechanism-equipped patterns mentioned in Table~\ref{t:rep}. For each pattern, we fabricate two tubes, one with reinforcing panels and one without, to compare their performance. Each tube features five unit cells along the axial direction and four along the circumference, and is capped by octagonal end caps. The cross section of tubes with panels is octagonal, while the cross section of tubes without reinforcing panel is circular, despite the octagonal end caps. Details on how the tubes are fabricated are reported in \ref{a:fab}.

\subsection{Morphing mechanics}

We begin by evaluating the morphing mechanics of tubes via experiments and numerical simulations. This analysis is needed to understand how ensembles of units behave, if they follow the mechanism motion predicted by kinematics, and to understand the role of the thickness --- a parameter that did not play a role in the kinematics but that is bound to have a prominent role on the behavior and mechanics of tubes.

Tubes are tested in tension using an universal testing system (UTS), by means of custom grippers that are anchored to the end caps of the tubes. Details on the experimental setup are reported in \ref{a:exp}. Tubes with reinforcing panels are also modeled via FE; in particular, we model the skin only, with boundary conditions that aim to replicate those of the experimental tubes, including the fact that bonding of the skin to the end caps will constrain the motion of the unit cells near the boundaries. Additional details on how the tubes are modeled, including the fact that we consider a strip of tube and apply periodic boundary conditions to it along the circumferential direction, are reported in \ref{a:num}.

Results of the morphing mechanics analysis are shown in Fig.~\ref{f:tubes} for the SN, SP, AN and AP patterns, respectively. These results include experimental force-displacement curves with and without stiffening panels and numerical curves for the stiffened cases only, in Figs.~\ref{f:tubes}(a)-(d). They also include snapshots of the deformation of tubes with (Figs.~\ref{f:tubes}(a-1)-(d-1)) and without stiffening panels (Figs.~\ref{f:tubes}(a-2)-(d-2)), together with numerical snapshots of the deformation of tubes with panels (Figs.~\ref{f:tubes}(a-3)-(d-3)).

For all designs, we can observe that the curves for tubes with panels are much more regular than those of tubes without them. In particular, we can see that the responses of tubes without panels for non-auxetic patterns feature more jagged responses than for the auxetic ones. The difference between cases with and without panels can be explained by looking at the experimental snapshots of the tube deformation. We can see that the tubes with panels follow a mode of deformation that resembles the mechanism, where panels rotate about each other and the folding hinges pop up. As a side note, a few hinges in each tube tend to pop-down towards the tube's center. The tubes without stiffening panels, conversely, do not follow the mechanism motion and are instead characterized by widespread localized buckling as in classical kirigami systems~\cite{rafsanjani2017buckling} --- which causes the jaggedness of the response. In other words, without stiffening panels, the skins almost behave as the fold lines are not present. This highlights the important of stiffening panels to obtain tubes that behave as desired and that follow a motion featuring counter-rotation of panels and pop-up of folding hinges.

The numerical simulations show that the stiffening panels bear the same regularizing effect as in the experiments, with deflected shapes indicating mechanism-like motion. From the deflected shapes, we can also appreciate how the cross sections towards the tube's midpoint shrink for non-auxetic designs and bulge out for auxetic designs, as expected. Overall, numerical and experimental curves show similar trends, but do not exactly overlap. This is due to the fact that, for simplicity, we add three connector elements per fold line, independently on the length of the fold line; while this is a good assumption if all fold lines have the same length, as in the SN pattern, this is not the case for the other patterns. As expected, the overlap is excellent for the SN pattern in Fig.~\ref{f:tubes}(a). The results for the other patterns, in Figs.~\ref{f:tubes}(b)-(d), show that the numerical response features an initial steep slope before flattening, as expected due to the buckling-like nature of the hinge pop-up. Such steep slope is not expected in experiments due to inevitable imperfections. It is expected that eventually the numerical curve should overlap with the experimental one at large displacements. While this is not visible for the limited range of simulated displacements in Figs.~\ref{f:tubes}(b), (c), it is clearly the case in Fig.~\ref{f:tubes}(d). We believe that further tuning of the numerical model (including adding connector elements to the fold lines according to a set density per unit length, and tuning the stiffness of the connector elements) would yield a better match with the experiments. To us, the fact that experiments and simulations results show similar trends despite the approximations in our models indicates that the response of tubes with stiffening panels is still dominated by the mechanism motion of the unit cells despite the presence of boundary conditions.

\begin{figure*}[!htb]
\centering
\includegraphics[scale=1]{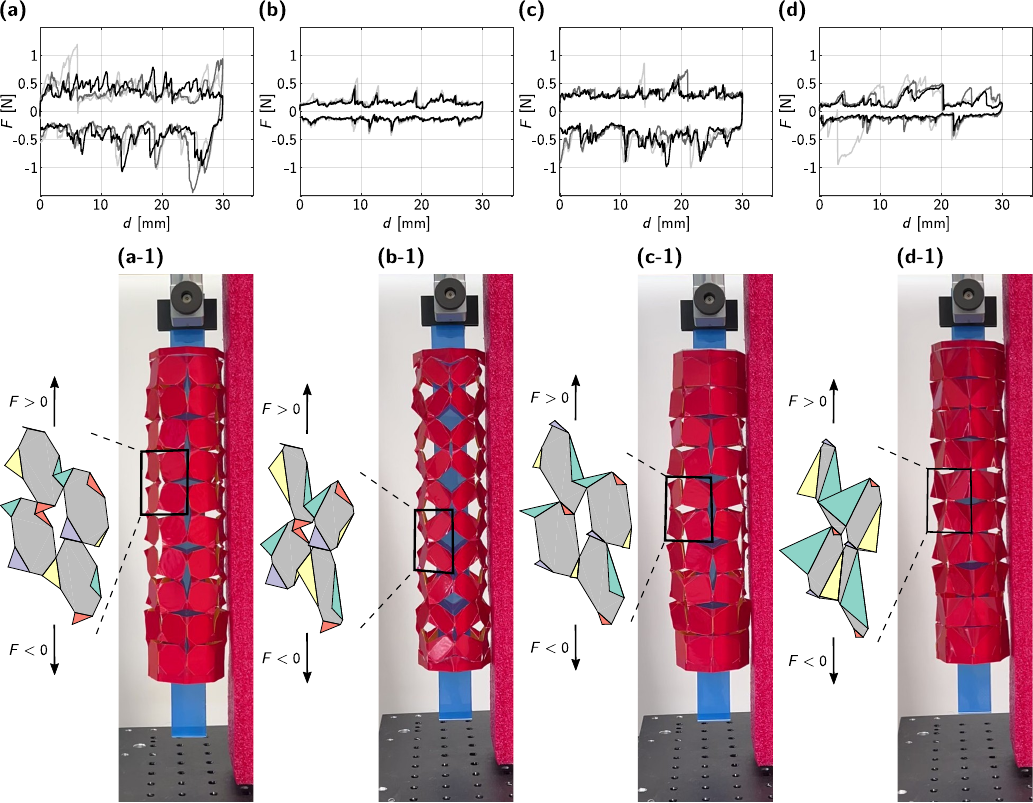}
\caption{Experimental results on frictional anisotropy, obtained by dragging the tubular specimens with stiffening panels along a rough polyethylene foam surface. The geometrical designs of the panels are labeled as in Table~\ref{t:rep}. (a) Response of the SN design, symmetric with negative Poisson's ratio. (b) Response of the SP design, symmetric with positive Poisson's ratio. (c) Response of the AN design, asymmetric with negative Poisson's ratio. (d) Response of the AP design, asymmetric with positive Poisson's ratio. In each plot, the three lines indicate three trials of the same experiment. (a-1)-(d-1) Photos of the tubes on the experimental setup, with legends indicating how the unit cell of each tubes is oriented with respect to the up ($F>0$) and down ($F<0$) loading directions.}
\label{f:frict}
\end{figure*}

\subsection{Preliminary results on anisotropic friction}

We conclude our study by investigating the frictional properties of our tubes. This investigation is performed via experiments only, on tubes that feature reinforcing panels. We first open up the tubes at a desired level of axial displacement via an actuating rod. Then, using our UTS, we drag the opened tubes along a rough polyethylene surface. For each experiment, we perform up-down motion cycles to try and unveil if the frictional forces show some anisotropy between upward and downward motion. Details on the experimental setup for these frictional tests are also reported in \ref{a:exp}.

The results of our frictional tests are shown in Fig.~\ref{f:frict}, for the same tubes tested in the previous section. Figs.~\ref{f:frict}(a)-(d) show the data measured by the UTS in terms of force-displacement curves, while Figs.~\ref{f:frict}(a-1)-(d-1), for each tube, indicate how the unit cells of the pattern are oriented with respect to loading ($F>0$) and unloading ($F<0$) directions. We begin by providing general observations on the force-displacement curves. The peaks/drops in the response indicate stick-slip, which is the main frictional mechanism when kirigami skins interact with surfaces~\cite{rafsanjani2018kirigami}. We can see that the occurrence of stick-slip is much more frequent in auxetic patterns (Figs.~\ref{f:frict}(a), (c)); this is not surprising, since these tubes bulge out when expanded, and therefore produce more normal force against the surface than their positive Poisson's ratio counterpart, if kept at a fixed distance from the surface. The outward bulging is also the reason why the average frictional force is larger for auxetic patterns. 

We now concentrate on the differences between symmetric and asymmetric patterns. We can see that symmetric patterns, SN and SP, show comparable loading-unloading responses both in terms of average force and magnitude/frequency of the peaks (Figs.~\ref{f:frict}(a), (b)). Conversely, we can see that the loading and unloading curves look different for the asymmetric patterns of Figs.~\ref{f:frict}(c), (d). In particular, from Fig.~\ref{f:frict}(c), we can see that the AN pattern presents a larger average force, higher and denser peaks during the unloading phase. This takes place when the small, thin yellow pop-ups interact with the surface, and it seems to highlight that smaller, stiffer pop-ups are more desirable to achieve high friction. Yet, the results of Fig.~\ref{f:frict}(d) for the AP pattern show a different story. In this case, the loading phase features larger average forces and more peaks, even though this corresponds to the larger (taller and with a wider area) green pop-ups interacting with the surface. 

In summary, our experimental results show that asymmetric patterns yield anisotropic friction; yet, this anisotropy is much smaller compared to other kirigami patterns~\cite{rafsanjani2018kirigami, rafsanjani2019propagation}, owing to the fact that our units always feature pop-ups that produce stick-slip behavior for both directions of motion rather than featuring smooth sliding along one direction of motion and stick-slip along the other. The limited data shown here is not enough for us to determine a relationship between the system's parameters, including pop-up height, size and Poisson's ratio, and the anisotropic frictional properties of these patterns. 

}

\section{Conclusions}
\label{s:conclusion}

In this work, we performed a comprehensive study on the morphing attributes of kirigami skins with folding hinges by means of analytical kinematics studies, numerical simulations and \textbf{red}{a range of experiments on unit cells and tubular skins. First, we showed that introducing asymmetry in these kirigami yields patterns with vastly-different responses in terms of global mechanisms and hinge pop-up behavior. Exploring this design space, we identified auxetic and non-auxetic designs with asymmetric pop-ups. We then showed, by studying the behavior of tubular skins, that stiffening panels are needed for the tubes to follow the mechanism-like motion of its unit cells. Finally, we provided a demonstration of frictional anisotropy in patterns with asymmetric pop-ups. As a result, we show the potential of these kirigami as skins} in the context of crawling and burrowing robots, or in other applications where asymmetric morphing or anisotropic friction are desirable.

 {Our results on anisotropic friction are preliminary: they showed anisotropy, but did not shed light on how the system's parameters affect friction. Yet, understanding this missing link and the interaction of pop-ups with rough surfaces or even granular media would be key to tune frictional properties and to the applicability of our metamaterials as skins for crawling and burrowing robots. We plan further studies in this direction, which we suspect will highlight that the parameter choices needed to achieve anisotropic friction in different media are different.}

Other extensions of this work, following analog studies for other kirigami systems~\cite{tang2019programmable, sempuku2021self, rafsanjani2019propagation}, could be related to: i) introducing asymmetric folding hinges in other planar kirigami patterns such as the kagome one; ii) studying the complete out-of-plane kinematics of these patterns, without constraining the large panels to remain in-plane; or iii) investigating how pop-ups propagate in these patterns and how to tune unit cell parameters across a tube to tailor the pop-up sequence.

\section*{Acknowledgments}

P.C. and A.K. acknowledge the support of the National Science Foundation (CMMI-2228271 and CMMI-2228272). A.W. acknowledges the support of Stony Brook University through a New York Space Grant research award. We thank Ahmad Rafsanjani from the University of Southern Denmark for sharing his Abaqus python codes, from which we learned how to apply periodic boundary conditions and simulate tubular patterns.

\section*{Author Contributions}

H.R.T.: Methodology; Software; Investigation; Writing - Original Draft. A.W.: Methodology; Investigation. A.K.: Conceptualization; Supervision; Funding acquisition; Writing - Review \& Editing. P.C.: Conceptualization; Methodology; Investigation; Visualization; Supervision; Funding acquisition; Writing - Original Draft.


\appendix


\section{Mechanism analysis}
\label{a:mechanisms}

 {To understand which combinations of parameters preserve global mechanisms,} we resort to linear matrix structural analysis arguments and we adopt the framework of Pellegrino and Calladine~\cite{pellegrino1986matrix}, which allows to determine mechanisms and states of self stress of truss assemblies, and whether the mechanisms are infinitesimal or finite in nature. As we will see, the presence of the folding hinges leads to additional challenges.

\begin{figure}[!htb]
\centering
\includegraphics[scale=1]{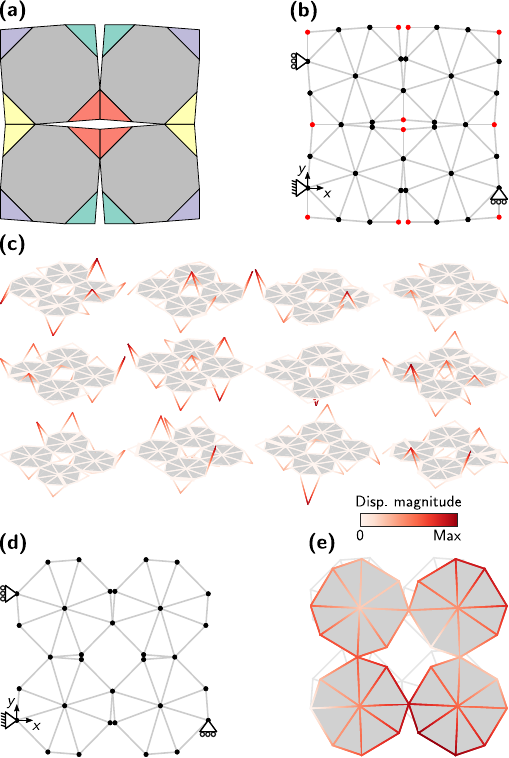}
\caption{Hierarchy of mechanisms. (a) Symmetric unit cell used to explain what mechanisms to expect in these systems. (b) Truss version of (a), where each octagonal panel has been made isostatic by connecting its boundary nodes to a newly created central node. Boundary conditions are as indicated in the figure; all black nodes are blocked along the out-of-plane direction $z$, while the red ones (tips of pop-ups) are not. (c) 12 ``local'' mechanisms of (b), resulting from our matrix analysis. (d) Version of (b) where pop-up triangles have been removed, assuming that their finite, zero-energy motions won't affect the global morphing of the unit cell. (e) ``Global'' mechanism of (d).}
\label{f:mechanisms}
\end{figure}

The first step in adopting this formalism is to transform our panel-based unit cells into trusses. One such transformation between a symmetric geometry and its truss analog is shown in Figs.~\ref{f:mechanisms}(a), (b). 
To model panels as rigid, it is not enough to place bars along their outer boundaries, as this would yield highly floppy constructs. There are many ways to triangulate octagons and making them rigid; we choose to add
a node to the center of each panel and connect all octagon nodes to the corresponding center node, as shown in Fig.~\ref{f:mechanisms}(b). Overall, the truss analog to our cell has $j=32$ joints and $b=64$ bars. In this truss structure, nodes can move out of plane; thus, the number of degrees of freedom per node is $n=3$. However, since we are interested in its planar morphing behavior (octagonal panels remain coplanar), we constrain all nodes to remain in the plane, apart from those at the tip of the pop-ups, shown in red in Fig.~\ref{f:mechanisms}(b). The planar boundary conditions we subject the unit to are shown in the same figure as pin and roller symbols. Importantly, these are chosen to be consistent with a mechanism motion that transforms an overall-rectangular cell to another rectangular shape --- a constraint that needs to be satisfied by a unit within a periodic pattern. 

The 12 mechanisms of the truss in Fig.~\ref{f:mechanisms}(b) are shown in Fig.~\ref{f:mechanisms}(c); by studying their interaction with the states of self stress, we determine that all these mechanisms are finite in nature. All of them are characterized by motions that are solely concentrated at the pop-ups, and there are as many mechanisms as the number of pop-ups. The fact that they are finite in nature indicates that these pop-ups are truly floppy and that, while they provide intriguing functionality to the metamaterial and affect its mechanics, they don't affect the motion of the octagonal panels. Notably, the desired global mechanism is missing from this set. We have a conjecture for its absence: the desired mechanism can only manifest \emph{after} the hinges pop-up; thus, there is a \emph{hierarchy} of mechanisms in this system, that linear analysis tools cannot identify. We therefore perform a second analysis, in which we remove the hinges and construct the hinge-less planar truss in Fig.~\ref{f:mechanisms}(d), where panels are connected at single points. We subject this structure to the same boundary conditions and study its mechanisms. We find that it has a finite mechanism consistent with the expected deformation of the system.

Now that we understand that folding hinges play a marginal role in the global kinematics of the unit, we perform a study on how the design parameters affect the presence/absence of the global mechanism. We are especially interested in understanding the role of the hinge lengths $\ell_i$ and the hinge angles $\lambda_i$. For this reason, we set $\bar{w}_x=\bar{w}_y=0$, confident that our findings can then be extended to other values of those parameters --- which are known not to affect the global mechanism in similar metamaterials~\cite{celli2018shape}. For each design, we follow these steps: i) create a truss version of the unit; ii) check if the truss has finite mechanisms and if these only involve hinge pop-up; if so, continue; iii) consider only the octagonal panels and create a truss version of this simplified unit; and iv) check if the truss has finite, infinitesimal or no global mechanism. 

\begin{figure*}[!htb]
\centering
\includegraphics[scale=1]{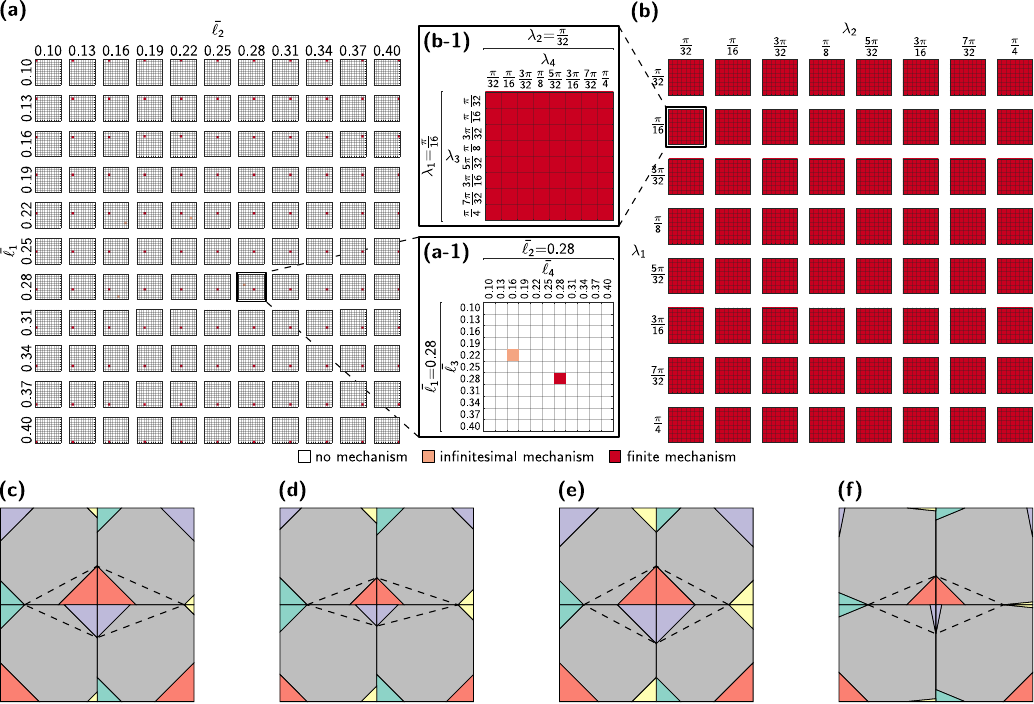}
\caption{Classification of asymmetric geometries based on their mechanism response, and on whether they display finite, infinitesimal or no global mechanism. (a) Colormap obtained by setting slit openings $\bar{w}_x=\bar{w}_y=0$, hinge angles $\lambda_1=\lambda_2=\lambda_3=\lambda_4=\pi/4$, and by varying hinge lengths $\bar{\ell}_1$, $\bar{\ell}_2$, $\bar{\ell}_3$ and $\bar{\ell}_4$. Each box in the plot corresponds to a combination of $\bar{\ell}_1$ and $\bar{\ell}_2$; in turn, each square within each box corresponds to a combination of $\bar{\ell}_3$ and $\bar{\ell}_4$, as shown in detail (a-1), which is an enlarged view of a single box of (a). (b) Colormap similar to the one in (a), where $\lambda_1$, $\lambda_2$, $\lambda_3$, $\lambda_4$ vary and we fix $\bar{w}_x=\bar{w}_y=0$ and $\bar{\ell}_1=\bar{\ell}_2=\bar{\ell}_3=\bar{\ell}_4=0.3$. (b-1) enlarged view of a box of (b). (c)--(f) Some of the geometries in our design space. In all of them, the dashed line connects the rotation points, and is akin to a slit in classical rotating squares kirigami. (c) Design that has no mechanism, from (a), with $\bar{\ell}_1=0.40$, $\bar{\ell}_2=0.25$, $\bar{\ell}_3=0.34$ and $\bar{\ell}_4=0.10$. (d) Design that has an infinitesimal mechanism, from (a), with $\bar{\ell}_1=0.18$, $\bar{\ell}_2=0.28$, $\bar{\ell}_3=0.22$ and $\bar{\ell}_4=0.16$. (e) Design with finite mechanism, from (a), with $\bar{\ell}_1=\bar{\ell}_3=0.40$ and $\bar{\ell}_2=\bar{\ell}_4=0.25$. (f) Design with finite mechanism, from (b), with $\lambda_1=\pi/4$, $\lambda_2=\pi/8$, $\lambda_3=\pi/16$, $\lambda_4=\pi/32$.}
\label{f:mechparam}
\end{figure*}

The results of this parametric analysis are shown in Fig.~\ref{f:mechparam}. In the first study, we set $\lambda_1=\lambda_2=\lambda_3=\lambda_4=\pi/4$ and investigate the effects of the individual hinge lengths. To do so, we use a nested colormap plot where, for each combination of $\bar{\ell}_1$ and $\bar{\ell}_2$ in a desired range, we plot a colormap as a function of $\bar{\ell}_3$ and $\bar{\ell}_4$, as shown in Fig.~\ref{f:mechparam}(a). A detail of a colormap for specific values of $\bar{\ell}_1$ and $\bar{\ell}_2$ is shown in inset (a-1). Points of the colormap are marked dark red if that combination of parameters yields a global mechanism of finite nature, pink if it yields an infinitesimal mechanism and white if it does not feature a mechanism. The range of values chosen for the normalized hinge lengths (0.1--0.4) avoids panel overlaps in the pattern. This figure shows that, to have a finite mechanism, we need always need $\bar{\ell}_1=\bar{\ell}_3$ and $\bar{\ell}_2=\bar{\ell}_4$. In other words, hinges across slits must have the same length. In the second study, we instead fix $\bar{\ell}_1=\bar{\ell}_2=\bar{\ell}_3=\bar{\ell}_4=0.3$ and we let the folding angles $\lambda_i$ vary. Again, we use a similar nested colormap to compactly present four-dimensional data; the results are shown in Fig.~\ref{f:mechparam}(b). We observe that, if all hinge lengths are identical, hinge angles can all be chosen independently of one another, as this does not affect the global mechanism. 

 {As mentioned in the main text, this finding echoes other studies which have pointed at the necessity of having parallelogram slits to preserve the global mechanism in rotating squares kirigami and their variations. To reconcile this comment with our specific kirigami}, we consider some of the geometries sampled form our design space, shown in Figs.~\ref{f:mechparam}(c)--(f), and add a dash line connecting the points about which the panels rotate. We can see that in Figs.~\ref{f:mechparam}(c), (d) --- designs that have all different values of hinge length and that yield infinitesimal or no mechanisms --- the line connecting the rotation points is not a parallelogram. On the other hand, in the designs of Figs.~\ref{f:mechparam}(e), (f) --- that feature finite mechanisms and where hinge lengths are either identical or such that opposite hinges have equal length, and where the angles are either identical or all different --- the dashed line forms a parallelogram {, confirming the affinity of our designs to classical rotating squares kirigami.}

\section{More on kinematics}
\label{a:kin}

In this appendix, we report on further steps needed to obtain the kinematic formulas discussed in Section~\ref{s:kinematics}, and discuss further limitations on the design parameters needed to avoid unwanted overlaps.

\subsection{Steps for the pop-up height formulas}

Here, we report some intermediate steps needed to obtain formulas for the pop-up heights $h_i$ as a function of $\theta$. We introduce some auxiliary parameters $a_i$ and $b_i$ for each pop-up, which define right triangles in which the third dimension is $\ell_i$ and one of the angles is $\lambda_i$, as shown in Fig.~\ref{f:desparam}(a) and as indicated in the following formulas:
\begin{equation}
        a_i=\frac{\ell_i}{\cos{\lambda_i}}, \qquad b_i=\ell_i \tan{\lambda_i}.
\end{equation}
To calculate the pop-up heights, we consider a deformed configuration of the unit cell (either Fig.~\ref{f:desparam}(b) or (c), where quantities of interest are only shown for the pop-up of the 1 type). We then concentrate on the triangle formed by the pop-up height $h_i$, the segment of length $b_i$, which is now also coming out of plane, and its projection on the $x$-$y$ plane. Using Pythagoras' theorem on this triangle, we determine that the height of a generic pop-up can be derived from
\begin{equation}
        b_i^2=h_i^2+\left(  \left( a_i \sin{(\lambda_i-\theta)}\right)^2 + \left( a_i \cos{(\lambda_i-\theta)} - \sqrt{\ell_i^2+h_i^2} \right)^2 \right).
    \label{e:heighteq}
\end{equation}
Solving this equation and keeping the positive solution yields
\begin{equation}
        h_i=\ell_i \sqrt{\frac{cos(2 \lambda_i - 2\theta) - cos(2 \lambda_i) }{cos(2\lambda_i - 2\theta) + 1}}.
\end{equation}

\subsection{Further restriction of the parameter space}

Some combinations of parameters yield unwanted overlaps between panels and hinges. To avoid overlaps, we set limits on these parameters, derived from geometrical considerations on the unit cell in Fig.~\ref{f:desparam}(a). Slit parameters $w_x$ and $w_y$ are chosen such that
\begin{equation}
    \begin{split}
        \begin{multlined}[t]
        w_x \leq \min \left(L_x-\ell_4-(\ell_1+w_y)\tan{\lambda_1},\right. \\
        \left.L_x-\ell_2-(\ell_3+w_y)\tan{\lambda_3}\right),
        \end{multlined}
        \\
        \begin{multlined}[t]
        w_y \leq \min \left( L_y-\ell_1-(\ell_2+w_x)\tan{\lambda_2},\right. \\ 
        \left.L_y-\ell_3-(\ell_4+w_x)\tan{\lambda_4} \right).
        \end{multlined}
    \end{split}
\end{equation}
Hinge lengths, on the other hand, are limited by 
\begin{equation}
    \begin{split}
        \ell_1 \leq L_y-w_y-(\ell_2+w_x)\tan \lambda_2,\\
        \ell_2 \leq L_x-w_x-(\ell_3+w_y)\tan \lambda_3,\\
        \ell_3 \leq L_y-w_y-(\ell_4+w_x)\tan \lambda_4,\\
        \ell_4 \leq L_x-w_x-(\ell_1+w_y)\tan \lambda_1.
    \end{split}
\end{equation}
Finally, hinge angles are limited by
\begin{equation}
    \begin{split}
        \lambda_1 \leq \arctan \left( \frac{L_x-\ell_4 -w_x}{\ell_1} \right), \quad \lambda_2 \leq \arctan \left( \frac{L_y-\ell_1 -w_y}{\ell_2} \right),\\
        \lambda_3 \leq \arctan \left( \frac{L_x-\ell_2 -w_x}{\ell_3} \right), \quad \lambda_4 \leq \arctan \left( \frac{L_y-\ell_3 -w_y}{\ell_4} \right).
    \end{split}
\end{equation}

\section{Fabrication details}
\label{a:fab}

{

Representative unit cells and tubular structures are fabricated via laser cutting and manual assembly. For all samples, we set the unit cell panel lengths to $L_x=L_y=2.5\,$cm. All samples feature a thin-sheet layer with fold lines and cuts, laser cut out of 0.05\,mm-thick PETG sheets. The cut pattern of one of our unit cells is shown in Fig.~\ref{f:fab}(a); to facilitate folding along desired fold lines, we weaken some regions by dotted cuts, with dots approximately 1\,mm distant from each other. These dotted cuts also act as stress relieving locations to prevent significant tearing at the fold lines (while some tearing after heavy usage is inevitable). 
\begin{figure}[!htb]
\centering
\includegraphics[scale=1]{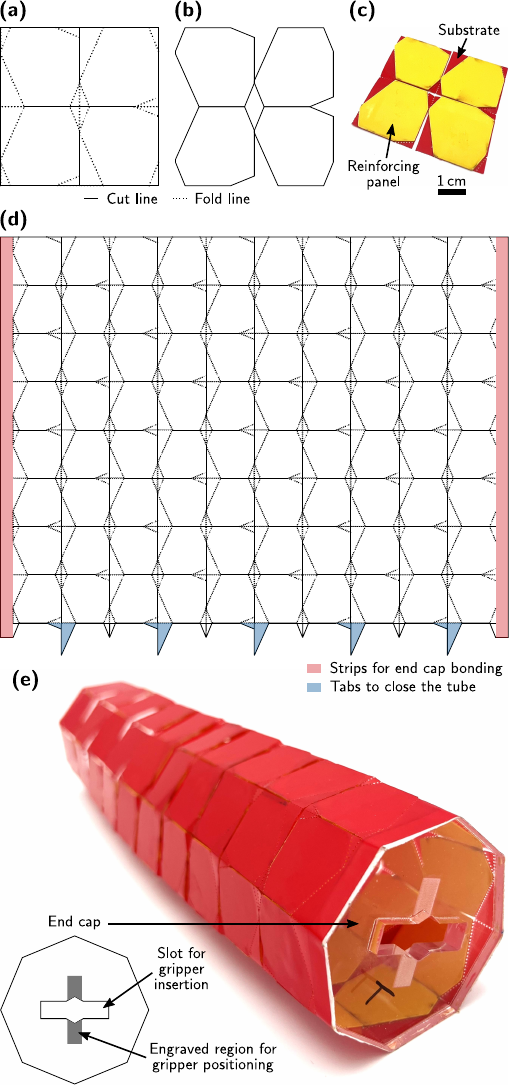}
\caption{Fabrication of unit cells and tubular structures. (a) Laser cut pattern for the thin substrate of a single unit cell of the AN type. (b) Laser cut pattern for the stiffening panels of the same unit cell; we directly laser cut panels with double-sided tape attached to them for easier assembly. (c) Assembled unit cell. (d) Laser cut pattern for the thin substrate of a tubular structure with AN geometry. The two vertical strips without cuts at the left and right edges are the locations where the octagonal end caps are bonded. The tabs on the bottom edge, where more double sided tape is located, are used to close the tube once the skin is wrapped around the end caps. (e) One of our fabricated tubes, with a drawing of one of the end caps.}
\label{f:fab}
\end{figure}

All of our unit cells and some of our tubular structures also feature stiffening panels, that allow the structures to closely follow the behavior expected from our kinematic calculations. To create these panels and to facilitate the process of bonding them to the thin substrate, we apply a double sided tape sheet (3M 300LSE) onto a 0.5\,mm-thick PETG sheet, and laser cut them together according to patterns as the one in Fig.~\ref{f:fab}(b). We then peel the other protective layer of the double sided tape and manually position the panels onto the substrate, obtaining unit cells as the one in Fig.~\ref{f:fab}(c).

Fabricating tubular structures requires some additional steps. Due to the presence of stiffening panels, our skins are only suitable to create tubes with polygonal, rather than circular, cross sections. Here, we choose to create tubes with octagonal sections, and this can be achieved by considering four cells along the circumferential direction. We instead choose five cells along the axial direction. We fabricate tubes of the four representative geometries in Table~\ref{t:rep}; for each geometry, we create a tube with stiffening panels and one with only the thin substrate, to compare their mechanics and the effects of these panels.

To facilitate bonding of octagonal end caps to our tube, we add strips without cuts to our laser cut pattern, as shown in Fig.~\ref{f:fab}(d). To be able to close the tubes once they are wrapped around the end caps, we add tabs to one of the two skin edges running along the axis of the tube, as shown in Fig.~\ref{f:fab}(d).

The tubes are assembled in two steps. First, we use double sided tape to bond the end caps to the skin. The end caps inevitably alter the way boundary unit cells behave, by constraining their deformation. The end caps have rectangular openings needed to fix them to our UTS via custom grippers, as discussed in \ref{a:exp}; it is important that the rectangular holes on the two end caps of the same tube are aligned with each other. The caps also feature an engraved rectangular area on their internal face, to easily position the custom gripper. After bonding the skin to the end caps, we use double sided tape to bond the tabs to the hinges on the opposite side of the skin, as to close the tubes without altering the kinematics of the unit cells at the bonded edge. A tube resulting from our fabrication process is shown in Fig.~\ref{f:fab}(e). 

}

\section{Experimental setups}
\label{a:exp}

 {In this appendix, we report details on our experimental procedures and the setups used throughout this work.}

\begin{figure}[!htb]
\centering
\includegraphics[scale=1]{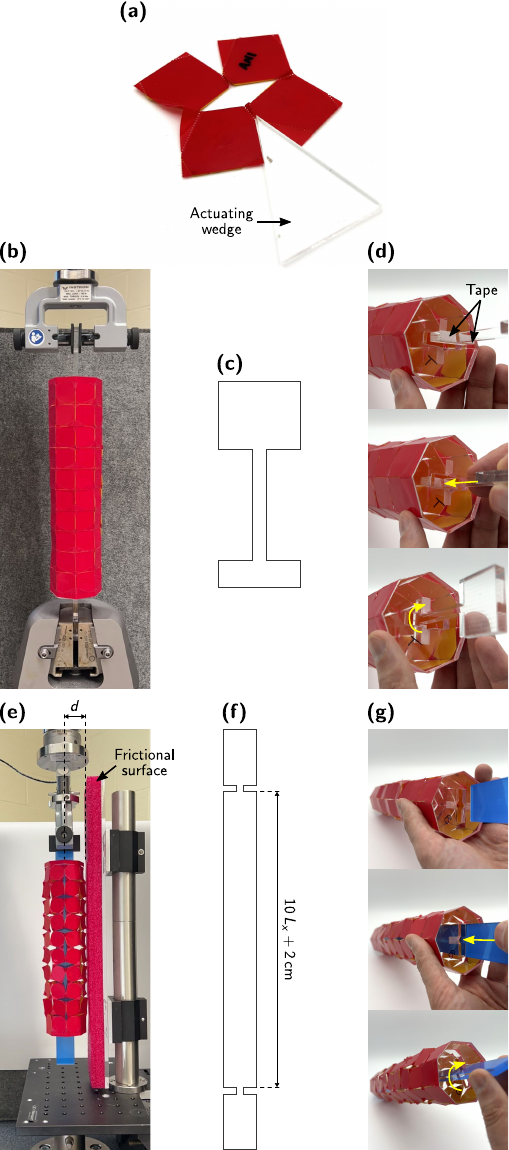}
\caption{Experimental setups. (a) Image of a unit cell, taped to a surface and opened at a desired angle $\theta=20^\text{o}$ via a wedge. Extension and pop-up heights are measured via a caliper. (b) Experimental setup to evaluate the morphing mechanics of tubes. (c) Drawing of the custom grippers. (d) Photos showing how one of the grippers is anchored to the tube via insertion, rotation and pulling. (e) Experimental setup to evaluate the anisotropic friction of a tube against a rough surface; $d$ is the distance between the axis of the tube and the surface. (f) Drawing of the actuation bar used to keep the tube at a desired displacement (2\,cm in this case) and to attach the open tube to the UTS. (g) Photos showing how the actuation bar is used to elongate the tube at a desired displacement.}
\label{f:exp}
\end{figure}

\subsection{Experiments to validate unit cell kinematics}

Our measurements on single unit cells are only intended to characterize how their shape changes during their mechanism motion, and therefore only probe their kinematics. To perform these measurements, we manually actuate the specimen to a desired opening angle with the aid of laser cut wedges, as shown in Fig.~\ref{f:exp}(a); if slit angles are nonzero to begin with, wedges account for the sum of the slit angle and the desired actuation angle. The octagonal panels are then taped to a flat surface with double-sided tape. We use a caliper to measure the overall cell extension (and, therefore, in-plane strains) and pop-up heights for various $\theta$ values. To account for specimen positioning and measurement inaccuracies, we perform each measurement three times, re-positioning the specimen each time. As a note, some plastic deformation and occasional tearing at the hinges is inevitable in these specimens. 

\subsection{Experiments on tube morphing mechanics}

 {The tubes are subjected to tensile loading on an Instron 68FM-100 Universal Testing System (UTS) equipped with a 100\,N load cell, at a loading rate of 1\,mm/s. A picture of how a tube specimen is loaded onto the machine is shown in Fig.~\ref{f:exp}(b). Grippers are laser cut out of 6.4\,mm-thick acrylic, and their shape is shown in Fig.~\ref{f:exp}(c). The top faces of the bottom anchor of the gripper are equipped with double sided tape. As shown in Fig.~\ref{f:exp}(d), the gripper is inserted into the end cap slot, then rotated and pulled once the bottom anchor is aligned with the engraved rectangle. Thanks to the tape, these grippers guarantee good adhesion to the tube end caps for tension experiments, at the low forces expected in our tests. For each tube, we carry out three measurements, from which we can compute an average and the standard deviation; the data is then reported as force-displacement curves in Fig.~\ref{f:tubes}.}

\subsection{Experiments on tube-surface friction}

 {To evaluate the frictional forces generated when tubes actuated at a desired displacement are dragged on a rough surface, we resort to the setup shown in Fig.~\ref{f:exp}(e). We manufacture an actuation bar, shown in Fig.~\ref{f:exp}(f), to keep the tube at a desired opening displacement, here chosen to be 2\,cm as this is the maximum displacement we subject our tubes to in the morphing mechanics experiments. This rod is inserted into the tube as shown in Fig.~\ref{f:exp}(g), and it is designed so it can be gripped by the top clamp of our UTS; the other end of the rod is kept free. To perform frictional test, we take a strip of polyethylene foam and bond it to a 6.4\,mm-thick acrylic plate. In turn, the plate is bonded to some fixtures to keep it vertical. The frictional surface is then placed at a distance $d$ from the axis of the tube, as shown in Fig.~\ref{f:exp}(e). This distance is chosen to be $d=3.2$\,mm, which is slightly larger than the maximum half diameter of the octagonal cross section of our tubes; our choice is such that the acrylic end caps do not touch the frictional surface during our test, and the measured frictional force is only a product of the interactions between skin and surface. To perform a test, we displace the tube up by 3\,cm and down by 3\,cm, at a rate of 1\,mm/s. When we test asymmetric skins, we expect the up ($F>0$) and down ($F<0$) tests to yield different results due to the asymmetry between pop-ups. Each test is repeated three times; each time, we reposition the frictional surface at the desired distance.}

\section{Numerical modeling details}
\label{a:num}

Here, we report some details on our numerical methods, and additional results that complement those in the main manuscript.

\subsection{General model information}

We model kirigami sheets in Abaqus with S4R5 shell elements; the connections between pop-ups and octagonal panels are modeled using connector elements~\cite{yang2023large}. The dimensions are chosen to be the same as those of the physical samples, with $L_x=L_y=2.5$\,cm. Panels are modeled with a linear elastic material model having Young's modulus of 2.2\,GPa and Poisson's ratio of 0.38, taken from measurements mentioned in Ref.~\cite{rahman2024shape}. The joint parameters are chosen to allow rotations about in-plane axes while restricting rotations about the out-of-plane axis; unless otherwise specified. The torsional stiffness about the in-plane and out-of-plane axes are chosen to be $k_{tx}=k_{ty}=10^{-6}$ and $k_{tz}=10^3$, respectively.  {Preliminary simulations suggest that a minimum of three hinges are necessary to properly connect each pop-up panel to its corresponding octagonal panel; thus, we place three connector elements along each fold line. As explained in the main text, this simplification is bound to create discrepancies between models and experiments.} 

\subsection{Unit cell models}

We first build unit cell models with periodic boundary conditions to simulate the behavior and kinematics of infinite sheets. The main challenge in implementing periodic boundary conditions for planar sheets is to correctly select boundary conditions to ensure stability and the application of the proper constraints on the unit cell. To do so, we basically set displacements of opposite fold lines to be equal and opposite (Fig.~\ref{f:geometry}(c) is useful to understand which fold lines need to be related to which in a prototypical pattern); this is done through two dummy nodes, i.e., reference points not connected to the cell but whose displacement is featured in the equations coupling the displacements of related fold lines. The first dummy node imposes essential boundary conditions and deformations from fold lines on the top edge to those on the bottom edge, while the second dummy node incorporates both essential and natural boundary conditions, aligning the right edge of the sheet with the left edge. Loading is then applied along the horizontal direction at the dummy node. 

In these unit cell simulations, panels are not constrained to remain in-plane, as these constraints cause convergence issues; it is observed that the panels can naturally remain almost completely planar. We use these periodic models to compare to the analytical and experimental results of a single cell, thus validating whether single cells are representative of periodic specimens. Since the pop-up of hinges is a buckling problem, we perform the analysis in two steps: first, we perform a linear buckling analysis and identify critical buckling loads and modes of interest, corresponding to the desired kinematics. These buckling modes are then used as imperfections in post-buckling investigations. Post-buckling analyses are performed in force-control mode with a quasi-static dynamic implicit method, with a sufficiently large time period chosen to minimize the effects of inertial forces.

\subsection{Tubular models}

 {We also built models for tubes with octagonal cross sections and stiffened panels. Exploiting the structural and loading symmetry observed in experiments, we only modeled a quarter of the geometry, with periodic boundary conditions enforcing the symmetry. We used a cylindrical coordinate system to define these conditions, which, similar to previous models, involved displacement constraints only. We then applied boundary conditions akin to those in the tensile test, reflecting the rpesence of end caps. Consequently, we used a single dummy node to define the displacement constraints along the symmetry lines, and we applied the tensile load directly to the model's nodes.}

 {Implementing boundary conditions for cylindrical and octagonal tubular patterns necessitates a distinct approach compared to planar kirigami. Notably, rotational constraints about the tube's axis are prioritized over displacement constraints between the top and bottom pop-ups of the unit cell. This ensures the preservation of overall symmetry and structural integrity. To further maintain a constant tube axis and tubular cross-section during deformation, we apply essential boundary conditions, specifically enforcing zero axial rotation of the octagonal panels within the unit cells.}

\subsection{On the effects of the hinge stiffness}

\begin{figure}[!htb]
\centering
\includegraphics[scale=1]{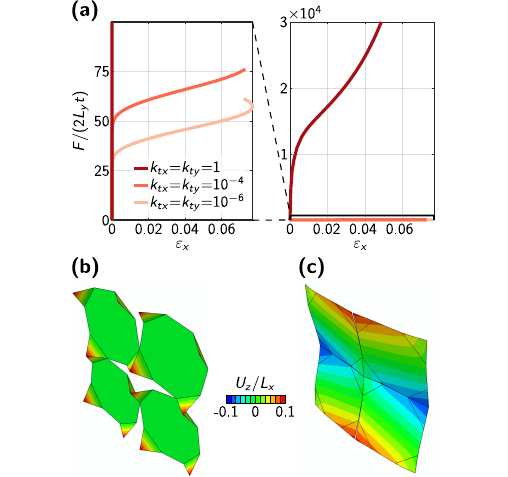}
\caption{Effects of hinge stiffness on the mechanics of single units. All graphs are stress-strain curves. (a) Mechanism loss due to increasing torsional stiffness about in-plane axes in the SN unit. The torsional stiffness about the out-of-plane $z$ axis is set to $10^3$. Snapshots of the deformation of that same pattern, with (b) $k_{tx}=k_{ty}=10^{-6}$ or $10^{-4}$, and (c) $k_{tx}=k_{ty}=1$. }
\label{f:mechanics}
\end{figure}
 {In our work, as previously mentioned, we choose specific values for the torsional hinge stiffness. Our choice is explained in the following.} To understand the effects of the in-plane torsional stiffness $k_{tx}$ (equal to $k_{ty}$) on the response of the pattern of Fig.~\ref{f:planar}(a-1), with periodic boundary conditions. The stress-strain curves for three different stiffness values are shown in Fig.~\ref{f:mechanics}(a), which includes a full-scale figure (right) and a zoom-in of it (left). Stress here is intended as the total applied force on the unit cell, normalized by the unit's cross sectional area ($2L_x t$). We can see that choosing a small $k_{tx}$ ($10^{-6}$ or $10^{-4}$) yields curves where the force rapidly increases and then plateaus as in Euler buckling problems, with the plateau here corresponding to the mechanism mode of deformation. Note that the curve for $10^{-6}$ curls back towards zero strain as boundary conditions in periodic patterns (essentially constraining regions along the boundary to behave the same way, without imposing displacements or forces on them) allow them to do so. While we can see that the energy needed to actuate these patterns varies with the torsional stiffness, the mechanism parts of the curve are very similar in slope, and the kinematics identical. For this reason, we choose $k_{tx}=k_{ty}=10^{-6}$ throughout this paper, as this value guarantees that the mechanism is present. Choosing larger values, e.g., $k_{tx}=k_{ty}=1$, yield instead a much stiffer response where the curve does not feature a plateau, as shown in the right panel of Fig.~\ref{f:mechanics}(a). The deformed unit cells in Fig.~\ref{f:mechanics}(b) and (c), corresponding to low and high $k_{tx}$, respectively, show indeed that the mechanism is lost when $k_{tx}=k_{ty}=1$ as highlighted by the overall bending of the unit cell and by the absence of pop-ups.

\end{document}